\documentclass{article}
\usepackage{spconf,amsmath,amssymb,graphicx,booktabs,multirow,array,url,hyperref,xcolor}
\usepackage{microtype}
\usepackage{enumitem}
\usepackage{dblfloatfix,placeins}
\usepackage{float}
\usepackage{balance}
\hypersetup{hidelinks,hypertexnames=false}

\newcommand{\vodercal}{VoDER-Cal}
\newcommand{\cosy}{CosyVoice3}
\newcommand{\vox}{VoxCPM2}
\newcommand{\fish}{Fish-Speech-S2}
\newcolumntype{L}[1]{>{\raggedright\arraybackslash}p{#1}}

\title{Beyond Prompt Adherence:\\Auditing Attribute-Level Voice Control in Speech Generation}
\name{Xianhao Zhou$^{1,\ast}$ \qquad Jianghao Wu$^{2}$}
\address{$^{1}$School of Mechanical and Electrical Engineering, UESTC, Chengdu, China\\
$^{2}$Faculty of Information Technology, Monash University, Melbourne, Australia\\
$^{\ast}$Corresponding author: \texttt{intelland2024@163.com}}

\begin{document}
\ninept
\maketitle

\begin{abstract}
Natural-language descriptions have become a flexible interface for controlling generated speech. Existing evaluations largely assess whether an output matches a prompt, but prompt matching alone does not reveal whether characteristics outside the intended change remain stable. We examine this distinction through a controlled paired audit of three speech-generation systems: \cosy{}, \vox{}, and \fish{}. The evaluation contains 5,940 outputs spanning six reference speakers, ten texts, three random seeds, and eleven conditions. Using acoustic, prosodic, content, and speaker measurements, we find that responses in the expected target direction are frequently accompanied by changes outside descriptor-specific signal-level target sets. This pattern remains among outputs whose target response exceeds baseline seed variation, and the accompanying changes differ substantially across systems. We further introduce \vodercal{}, a training-free candidate selector that retains sufficiently strong target responses while favoring smaller off-target deviations. A three-candidate pool raises the joint success rate from 4.8\% under single-sample direct generation to approximately 14\% for all candidate-selection policies. Within the matched three-candidate budget, \vodercal{} reduces held-out off-target deviation from 0.344 under target-only selection to 0.276 and improves listener-rated preservation. Preservation-sensitive evaluation therefore complements prompt-adherence evaluation, while candidate reranking offers a practical inference-time improvement. Code, configuration files, and analysis scripts are available at \url{https://github.com/intelland/VoDER}.
\end{abstract}


\begin{figure*}[t]
\centering
\includegraphics[width=0.99\textwidth]{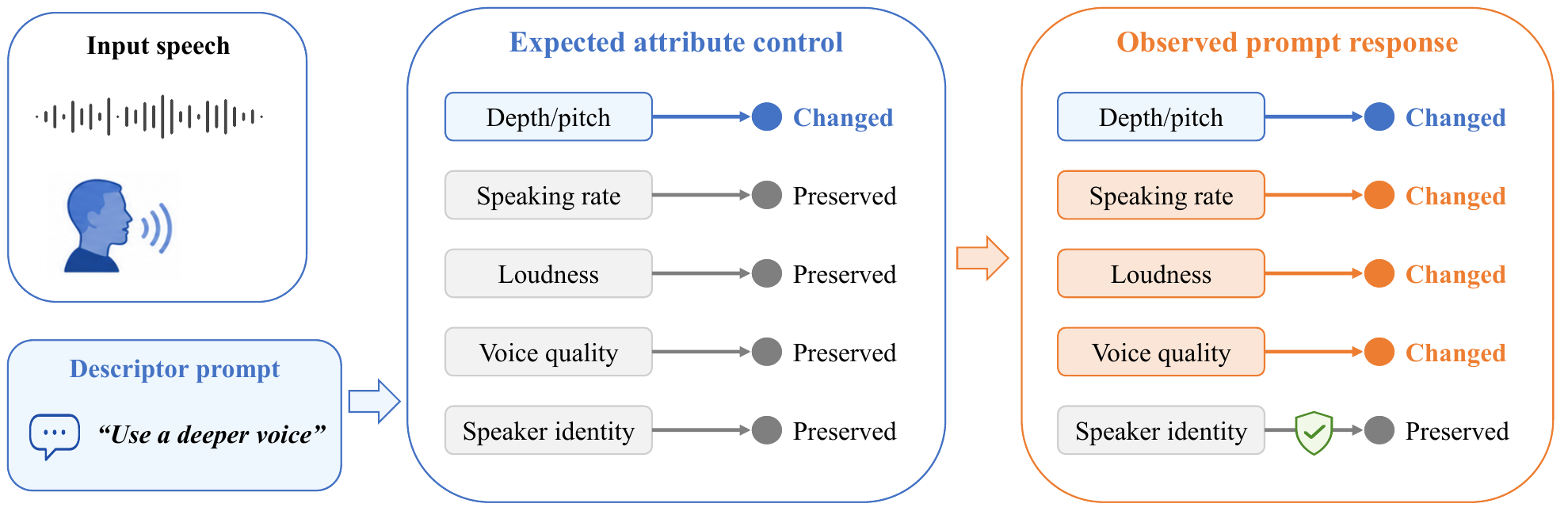}
\caption{Prompt response and a localized attribute edit answer different questions. The paired audit measures the requested change together with movement outside a descriptor-specific signal-level target set. \vodercal{} uses candidate variability to retain the target response while favoring smaller off-target movement.}
\label{fig:overview}
\end{figure*}

\section{Introduction}
\label{sec:intro}

Natural language allows users to control generated speech without relying on predefined style labels or low-level acoustic parameters. A desired voice can be described using attributes such as \emph{deep}, \emph{bright}, or \emph{rough}, or through more detailed free-form instructions. PromptTTS, InstructTTS, PromptTTS~2, PromptStyle, and PromptTTS++ demonstrated the feasibility of using text descriptions to control speaking style and voice characteristics \cite{guo2023prompttts,yang2024instructtts,leng2024prompttts2,liu2023promptstyle,shimizu2024promptttspp}. More recent systems combine natural-language control with reference conditioning, zero-shot voice generation, and open-vocabulary instructions \cite{ji2025controlspeech,du2025cosyvoice3,zhou2026voxcpm2,liao2026fishaudios2,ren2026ovinstructtts,chen2026flexivoice}.

Evaluation has largely focused on whether generated speech is compatible with the requested description. Common measures include prompt or style preference, naturalness, intelligibility, and speaker similarity, while recent benchmarks directly assess instruction-following fidelity \cite{huang2025instructttseval,cho2026spam}. These evaluations reveal whether a system responds to a prompt, but not whether it changes the requested attribute while preserving other characteristics of the voice. Prompt adherence and attribute-specific control are therefore related but distinct capabilities.

Consider a request for a deep voice. Lowering fundamental frequency is a plausible response, but the resulting speech may also become slower, quieter, spectrally darker, or less similar to the reference speaker. Such an output may sound more compatible with the word \emph{deep}, yet fail to provide a localized and reproducible edit. This distinction matters when users want to preserve a reference voice, combine several controls, or apply the same modification consistently across speakers and texts. We therefore ask:
\begin{quote}
When a system responds to a voice descriptor, what changes outside the requested attribute?
\end{quote}
We operationalize this question with descriptor-specific signal-level target sets defined before the final evaluation. Measurements outside the corresponding set are treated as off-target for the audit.

We study this question through paired generation. For the same system, reference speaker, text, and nominal random seed, we compare a neutral output with a descriptor-conditioned output. We evaluate three independently developed systems using a shared matrix of speakers, texts, seeds, descriptors, and controls. Acoustic and prosodic measurements quantify target and off-target movement, while ASR and speaker embeddings assess content and speaker preservation.

The results reveal a gap between descriptor response and preservation. Systems often move in the expected target direction, particularly for \emph{deep}, while simultaneously changing several non-target characteristics. The response profiles differ substantially across systems: the same descriptor may be realized through different combinations of pitch, timing, energy, and spectral change. The pattern persists after restricting the analysis to outputs whose target response exceeds baseline seed variation. An auxiliary listening study finds that both requested and accompanying non-target changes can be perceptually noticeable.

Candidate variability provides an opportunity to improve this behavior at inference time. Samples generated for the same request often express the target attribute with different amounts of collateral movement. We therefore introduce \vodercal{}, a training-free candidate selector that retains sufficiently strong target responses and ranks feasible candidates by off-target deviation. It requires neither generator retraining nor access to internal model representations. The candidate pool chiefly improves the availability of feasible outputs, while \vodercal{} improves preservation relative to target-only selection under the same generation budget.

Our contributions are:
\begin{itemize}[leftmargin=1.45em,itemsep=2pt,topsep=2pt]
\item a paired, preservation-sensitive audit based on matched neutral and descriptor-conditioned generations;
\item a cross-system study of 5,940 outputs with request-level statistical inference; and
\item \vodercal{}, a training-free candidate selector evaluated against single-sample, matched-budget, and oracle baselines.
\end{itemize}

\section{Related Work}
\label{sec:related}

\subsection{Natural-language control of speech}

Controllable speech generation has traditionally relied on either global style representations or explicitly defined acoustic factors. Reference-based prosody transfer and global style tokens encode speaking characteristics in learned latent representations \cite{skerry2018prosody,wang2018gst}, while architectures such as FastSpeech~2 expose duration, pitch, and energy as explicit variance controls \cite{ren2021fastspeech2}. These approaches reflect two different forms of control: matching a holistic style representation and manipulating an identifiable speech factor.

Natural-language prompting provides a more flexible alternative by allowing users to describe a desired speaking style or voice directly. PromptTTS maps textual descriptions to speaking styles \cite{guo2023prompttts}, while InstructTTS studies expressive instruction following and reduces leakage among style, speaker, and linguistic content \cite{yang2024instructtts}. PromptTTS~2 and PromptTTS++ further expand text-described voice variation and speaker control \cite{leng2024prompttts2,shimizu2024promptttspp}. ControlSpeech combines reference-based speaker cloning with language-guided style control \cite{ji2025controlspeech}. More recent systems extend natural-language control toward open-vocabulary instructions, zero-shot generation, and more structured speech representations \cite{ren2026ovinstructtts,chen2026flexivoice,lee2026fctts,li2025discospeech}. Our work studies a complementary question: when these systems respond to a voice description, how specifically is the requested attribute changed?

\subsection{Evaluating controllability}

Controllable TTS systems are commonly evaluated using prompt or style preference, naturalness, intelligibility, speaker similarity, and selected acoustic correlates. Recent benchmarks more directly assess whether generated speech follows complex natural-language instructions. InstructTTSEval organizes such instructions into a systematic evaluation benchmark \cite{huang2025instructttseval}, while SPAM develops a human-aligned measure of prompt adherence \cite{cho2026spam}.

These evaluations primarily measure semantic or perceptual agreement with the requested description, together with general generation quality and validity. They do not directly test whether characteristics outside the requested attribute remain stable relative to a matched baseline. Our paired audit adds this preservation-sensitive dimension by separating target response from off-target movement. The analysis is behavioral: it characterizes changes in generated outputs and does not infer whether the underlying models contain disentangled internal representations.

\subsection{Inference-time candidate selection}

Multi-candidate generation and reranking provide a training-free way to improve the reliability of speech generation. In zero-shot TTS, Best-of-$N$ inference generates several candidates and uses an ASR-based verifier to select the output whose recognized transcript most closely matches the target text \cite{yu2026bontts}. ASR self-verification has also been used to filter stochastic failures in neural-codec TTS, including silence, early termination, repetition, and hallucinated content \cite{asaria2026reliabletts}. These approaches demonstrate that useful variation among sampled candidates can be exploited at inference time.

\vodercal{} follows this multi-candidate selection paradigm but optimizes a different aspect of controllability. It first retains candidates that express the requested target response while preserving content and speaker identity, and then favors candidates with smaller movement in non-target acoustic and prosodic measurements. Thus, rather than selecting candidates only for content reliability, \vodercal{} uses inference-time variation to improve attribute preservation in natural-language voice control.

\section{Paired Evaluation of Attribute-Level Control}
\label{sec:audit_method}

\subsection{Systems and generation matrix}

We evaluate \cosy{}, \vox{}, and \fish{}, three independently developed reference-conditioned speech-generation systems. Each system is tested using the same factorial matrix of six reference speakers, ten English texts, three random seeds, and eleven conditions: one neutral baseline, six voice descriptors, and four controls. This yields 1,980 outputs per system and 5,940 outputs overall.

The primary cross-model audit and candidate-selection analysis focus on \emph{deep}, \emph{bright}, and \emph{rough}, for which the evaluation protocol defines directly comparable signal-level target proxies. The remaining descriptors---\emph{warm}, \emph{young}, and \emph{professional}---and the identity-preservation control are retained as supplementary audit conditions and are not part of the primary claims.

Each descriptor-conditioned output is paired with a neutral baseline generated by the same system using the same reference speaker, text, and nominal random seed. This design allows descriptor-induced changes to be measured relative to a closely matched baseline. We retain the native control interface of each system; in particular, \fish{} uses its inline voice-tag syntax rather than a free-form instruction channel. Model identifiers and control interfaces are provided in the supplementary material.

\subsection{Paired feature analysis}

Let $y_0$ denote a neutral baseline output and $y_d$ the paired output conditioned on descriptor $d$. For acoustic or prosodic feature $z_j$, the paired change is
\begin{equation}
\Delta z_j=z_j(y_d)-z_j(y_0).
\end{equation}

We measure pitch, timing, energy, spectral color, and signal irregularity using F0, speaking rate, duration, RMS energy, spectral centroid, 85\% spectral roll-off, spectral flatness, and zero-crossing rate. F0 changes are expressed in semitones. These are standard signal-level measurements rather than outputs of a learned attribute evaluator \cite{mcfree2015librosa,mauch2014pyin}.

For each descriptor, we define a signal-level target set and its expected direction:
\begin{align}
\textit{deep}:   &\quad \mathrm{F0}\downarrow,\ \mathrm{rolloff}\downarrow,\\
\textit{bright}: &\quad \mathrm{centroid}\uparrow,\ \mathrm{rolloff}\uparrow,\\
\textit{rough}:  &\quad \mathrm{flatness}\uparrow,\ \mathrm{ZCR}\uparrow.
\end{align}

Features outside the corresponding target set are treated as off-target measurements. These sets define the scope of the signal-level audit rather than exhaustive perceptual models of the descriptors. Spectral flatness and zero-crossing rate are limited proxies for roughness.

To compare movements across features, we normalize absolute changes using fixed, physically interpretable scales shared across all systems. Because several measurements are correlated, counts of shifted features describe the breadth of the measured response, not the number of independent perceptual attributes. We therefore report both individual movements and grouped summaries.

\subsection{Validity checks and statistical inference}

We assess content preservation using normalized ASR transcripts and speaker preservation using cosine similarity between generated-speech and reference-speech ECAPA-TDNN embeddings; x-vector similarity provides an additional cross-check \cite{radford2023whisper,desplanques2020ecapa,snyder2018xvectors}. Binary content and speaker thresholds are determined on the calibration split and then applied unchanged to the evaluation set. These measures serve as validity checks for excluding cases dominated by content corruption or speaker replacement.

Statistical inference is performed at the request level, where a request is defined by a system, descriptor, reference speaker, and text. For mean feature effects, the three seed-level deltas are first averaged within each request, and request blocks are then resampled with replacement. This avoids treating repeated generations of the same request as independent observations. Confidence intervals use 10,000 bootstrap replicates.

Conditional-response proportions are computed at the output level, but uncertainty is estimated by resampling request clusters. Outputs sharing the same system, descriptor, speaker, and text are therefore kept together during resampling.

\subsection{Conditional analysis and controls}

To distinguish collateral movement from failures to express the descriptor, we perform a second analysis conditioned on a clear target response. Baseline variation across random seeds defines a one-sided target-direction noise threshold for each descriptor. An output is considered target-responsive when its target score exceeds this threshold. Among these outputs, we measure the fraction exhibiting at least one off-target change whose magnitude reaches half of its fixed normalization scale. Request-level bootstrap intervals are also used to identify off-target features with systematic nonzero movement.

The control set includes a neutral reading instruction, a nonsense descriptor, an identity-preservation instruction, and an explicit lower-pitch instruction. The lower-pitch condition provides a positive control for the generation interface and F0 measurement. The neutral and nonsense conditions quantify perturbations introduced by additional natural-language conditioning in the absence of a meaningful target voice attribute.

\section{Voice Descriptors Produce Coupled Changes}
\label{sec:audit_results}

\subsection{Models respond, but not through isolated changes}

\begin{figure*}[t]
\centering
\includegraphics[width=0.99\textwidth]{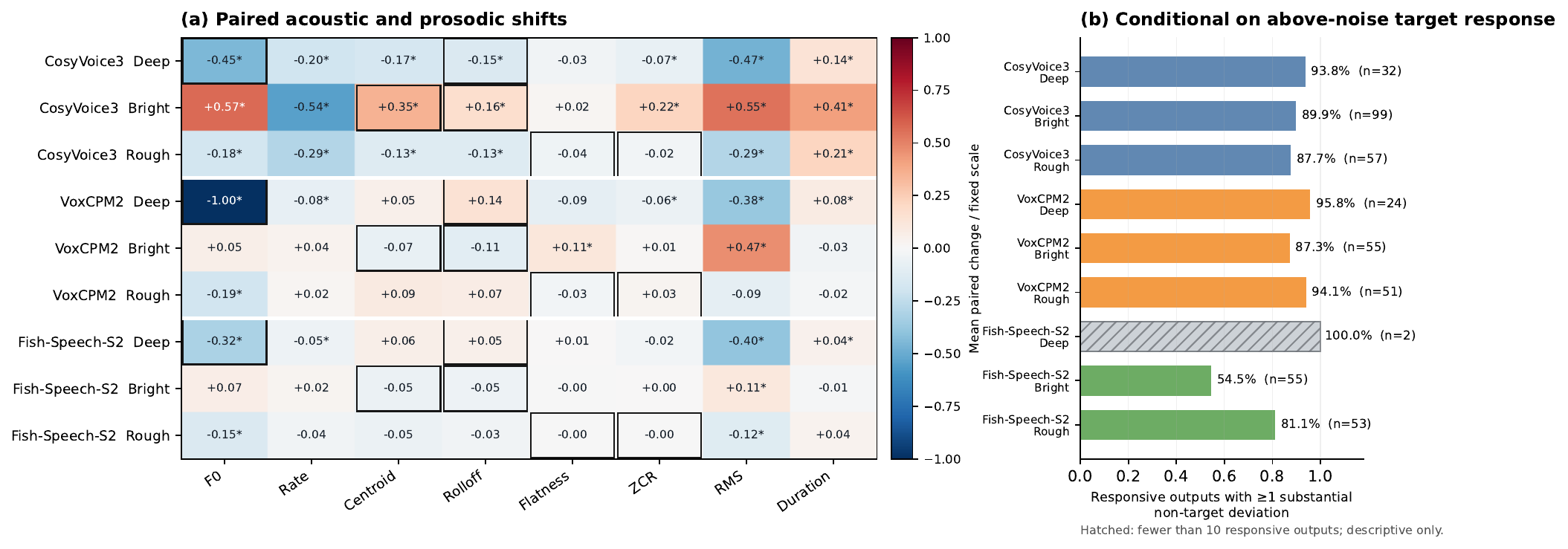}
\caption{Descriptor-induced changes. (a) Mean paired feature shifts normalized by fixed physical scales. Black boxes mark target features, and asterisks indicate request-level 95\% bootstrap confidence intervals excluding zero. (b) Among outputs whose target response exceeds baseline seed variation, the fraction exhibiting at least one substantial off-target deviation. The \fish{} deep result is based on only two target-responsive outputs and is therefore descriptive.}
\label{fig:audit}
\end{figure*}

Figure~\ref{fig:audit}a shows that the three systems respond systematically to the tested descriptors, but their responses frequently extend beyond the corresponding target features. For \emph{deep}, outputs move in the expected target direction in 84.4\% of \cosy{} pairs, 77.2\% of \vox{} pairs, and 71.1\% of \fish{} pairs. Responses to \emph{bright} and \emph{rough} vary more across systems and measured features.

Target-aligned changes are rarely isolated. For example, \cosy{} lowers F0 in response to \emph{deep}, while also changing speaking rate, duration, RMS energy, spectral centroid, and roll-off. Its \emph{bright} response increases the target spectral measures but also changes speaking rate and energy. \vox{} produces a strong F0 reduction for \emph{deep} together with lower RMS energy, whereas its \emph{bright} response is expressed more strongly through energy than through the expected spectral targets. Overall, the same descriptor is realized through different combinations of pitch, timing, energy, and spectral change across systems.

Table~\ref{tab:audit} combines target responsiveness with content, speaker, and off-target preservation measures. Content is preserved in at least 99.4\% of outputs, and mean reference-speaker similarity ranges from 0.821 to 0.902. The coupled acoustic changes therefore occur largely among outputs that retain both linguistic content and reference-speaker identity. Multiple off-target features show request-level confidence intervals excluding zero in most system--descriptor settings. These counts summarize the breadth of measured change.

\begin{table*}[t]
\centering
\caption{Primary paired audit. ``Target dir.'' is the fraction of
outputs moving in the expected target direction. ``Non-target CIs''
counts non-target features whose request-level confidence intervals
exclude zero. ``Conditional'' is the fraction of above-noise
target-responsive outputs with at least one substantial off-target
deviation.}
\label{tab:audit}
\resizebox{0.94\textwidth}{!}{\begin{tabular}{llrrrrr}
\toprule
Model & Descriptor & Target dir. & Non-target CIs & Conditional & Content & Spk. sim.\\
\midrule
CosyVoice3 & Deep & 84.4\% & 5/6 & 93.8\% (n=32) & 100.0\% & 0.886\\
 & Bright & 71.1\% & 5/6 & 89.9\% (n=99) & 100.0\% & 0.902\\
 & Rough$^{\dagger}$ & 44.4\% & 6/6 & 87.7\% (n=57) & 100.0\% & 0.890\\
\addlinespace[1pt]
VoxCPM2 & Deep & 77.2\% & 4/6 & 95.8\% (n=24) & 99.4\% & 0.821\\
 & Bright & 47.8\% & 2/6 & 87.3\% (n=55) & 100.0\% & 0.846\\
 & Rough$^{\dagger}$ & 53.9\% & 1/6 & 94.1\% (n=51) & 100.0\% & 0.845\\
\addlinespace[1pt]
Fish-Speech-S2 & Deep & 71.1\% & 3/6 & 100.0\% (n=2)$^{\ddagger}$ & 100.0\% & 0.895\\
 & Bright & 48.9\% & 1/6 & 54.5\% (n=55) & 100.0\% & 0.897\\
 & Rough$^{\dagger}$ & 49.4\% & 2/6 & 81.1\% (n=53) & 100.0\% & 0.898\\
\bottomrule
\end{tabular}
}
\vspace{-2pt}
\begin{flushleft}\footnotesize
$^{\dagger}$ Flatness and zero-crossing rate are weak automatic proxies for roughness. $^{\ddagger}$ Fewer than ten target-responsive outputs; descriptive only.
\end{flushleft}
\end{table*}

\subsection{Off-target changes remain after conditioning on response}

Conditioning on a clear target response does not remove off-target change. Across the eight system--descriptor settings with at least ten above-noise target-responsive outputs, 54.5\%--95.8\% of responsive outputs exhibit at least one substantial off-target deviation (Fig.~\ref{fig:audit}b). Rates reach 93.8\% and 89.9\% for \cosy{} \emph{deep} and \emph{bright}, and 95.8\%, 87.3\%, and 94.1\% for \vox{} \emph{deep}, \emph{bright}, and \emph{rough}. Even the lowest sufficiently populated setting, \fish{} \emph{bright}, reaches 54.5\%. Across these settings, two to five off-target features also show systematic request-level effects with confidence intervals excluding zero.

Off-target movement is therefore not confined to samples that fail to express the descriptor. Outputs can exhibit an above-noise target response while simultaneously changing other acoustic or prosodic characteristics. A clear target response alone is not sufficient evidence of an attribute-specific edit.

\subsection{Controls and robustness}

The explicit lower-pitch control produces the expected mean F0 decrease in all three systems: $-7.33$ Hz for \cosy{}, $-7.17$ Hz for \vox{}, and $-4.80$ Hz for \fish{}, with all bootstrap confidence intervals below zero. This confirms that the generation interfaces and F0 measurement respond to a direct low-level intervention.

Neutral and nonsense conditions also shift several measurements in some systems. Additional natural-language conditioning can therefore perturb the generated voice even without a meaningful target descriptor. These controls further motivate evaluating descriptor-conditioned speech relative to a matched baseline rather than interpreting conditioned outputs in isolation. Detailed control contrasts are reported in the supplementary material.

\section{VoDER-Cal: Preservation-Aware Candidate Selection}
\label{sec:voder}

The paired audit reveals substantial variation among candidates generated for the same request. Different candidates can achieve similar target responses while exhibiting markedly different off-target changes. \vodercal{} exploits this variation to improve attribute preservation without retraining the generator or accessing its internal representations.

\subsection{Candidate selection}

For each request, we generate a bounded candidate set $\{y_i\}_{i=1}^{B}$ and compare every candidate with the matched neutral baseline $y_0$. Let $S_d$ denote the target feature set for descriptor $d$, $a_j$ the normalization scale for feature $j$, and $s_{d,j}\in\{-1,+1\}$ its expected direction. Candidate $i$ receives the direction-aligned target score
\begin{equation}
T_i(d)=\frac{1}{|S_d|}
\sum_{j\in S_d}
s_{d,j}\frac{z_j(y_i)-z_j(y_0)}{a_j}.
\label{eq:target_score}
\end{equation}
This score averages normalized target-aligned changes across the descriptor's target features.

Let $C_i$ and $I_i$ denote the content- and speaker-validity checks. The feasible candidate set is
\begin{equation}
\mathcal{F}=
\left\{
y_i:
T_i\geq
\max(\tau_{\mathrm{noise}},\rho T_{\max}),
\ C_i=1,\ I_i=1
\right\},
\label{eq:feasible}
\end{equation}
where $\tau_{\mathrm{noise}}$ is the target-response threshold estimated from baseline seed variation, $T_{\max}$ is the strongest target score in the candidate pool, and $\rho$ controls target retention. A candidate is feasible if it exceeds baseline variation, retains at least a fraction $\rho$ of the strongest available target response, and preserves content and speaker identity.

For each feasible candidate, we compute the mean absolute normalized change over a selector-side set of non-target features, denoted by $D_{\mathrm{off}}^{\mathrm{select}}(y_i,y_0)$. \vodercal{} selects
\begin{equation}
y^{\star}
=
\arg\min_{y_i\in\mathcal{F}}
D_{\mathrm{off}}^{\mathrm{select}}(y_i,y_0).
\label{eq:select}
\end{equation}
If no candidate is feasible, the method abstains. To evaluate preservation beyond the measurements used for ranking, automatic evaluation uses a non-overlapping set of held-out off-target features. The main operating point uses $B=3$ and $\rho=0.75$. Feature partitions, thresholds, and normalization scales are determined on the calibration split and then applied unchanged to evaluation requests. Alternative selector--evaluator partitions and calibration-only robust scaling are reported in the supplementary material.

\subsection{Automatic evaluation}

We compare five inference policies. Direct uses a single fixed-seed generation, and Random uses one randomly selected generation. Target-only, \vodercal{}, and the candidate-pool Oracle select from the same three-candidate pool. Target-only chooses the feasible candidate with the strongest target response, while the Oracle chooses the candidate with the lowest held-out off-target deviation. The generation budget counts descriptor-conditioned candidates; the matched neutral baseline is shared across policies. A request is counted as a joint success when the selected output satisfies the target, content, speaker, and held-out preservation criteria.

\begin{figure*}[t]
    \centering

    \begin{minipage}[c]{0.48\textwidth}
        \centering

        {\footnotesize\bfseries
        (a) \vodercal{}
        \par}

        \vspace{1mm}

        \includegraphics[
            width=0.92\linewidth
        ]{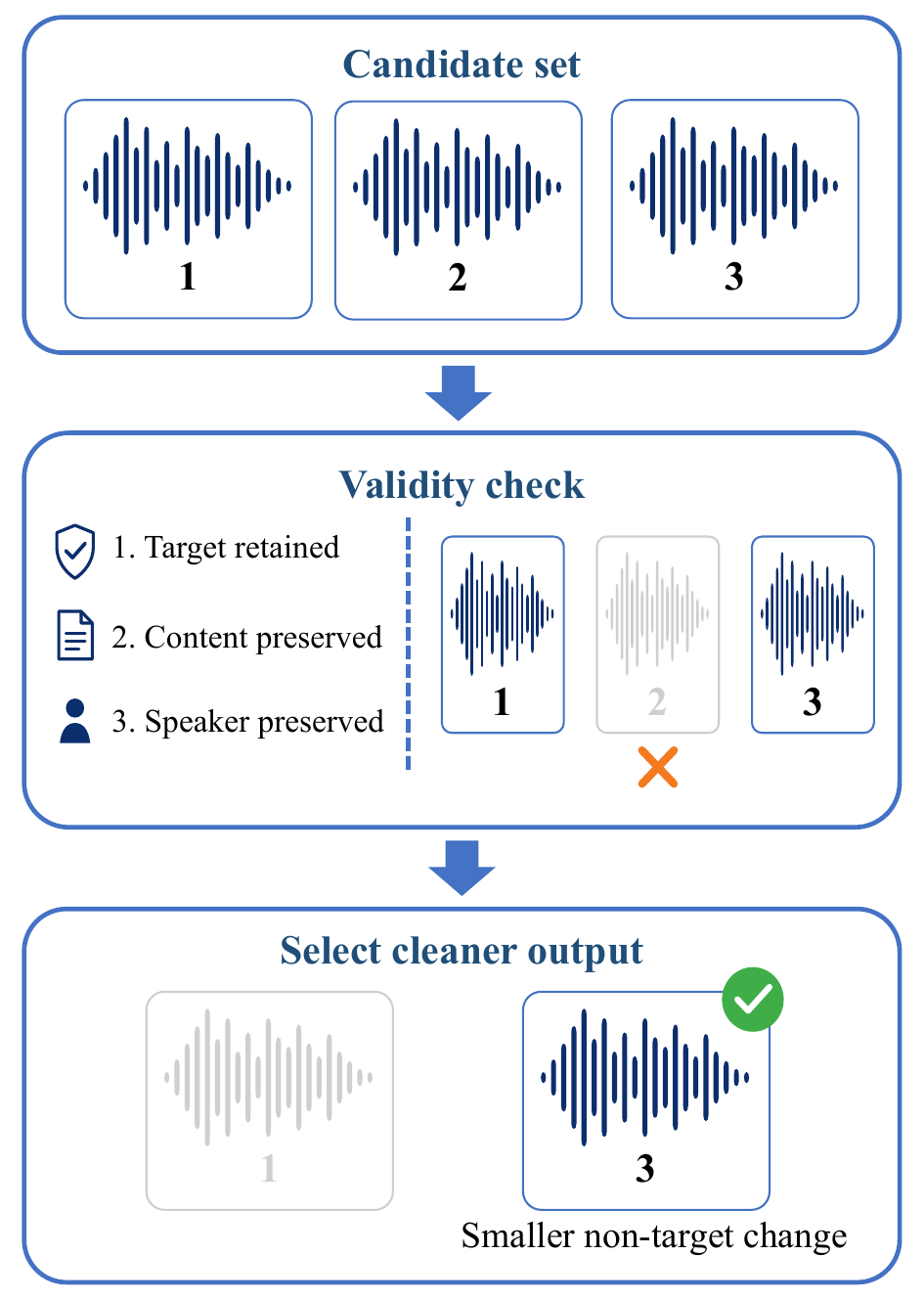}
    \end{minipage}
    \hfill
    \begin{minipage}[c]{0.48\textwidth}
        \centering

        {\footnotesize\bfseries
        (b) Joint success rate
        \par}

        \vspace{0.5mm}

        \includegraphics[
            width=0.90\linewidth
        ]{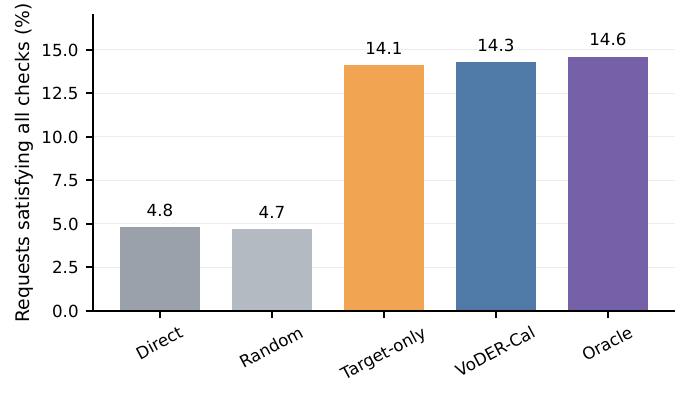}

        \vspace{2.5mm}

        {\footnotesize\bfseries
        (c) Candidate ranking
        \par}

        \vspace{0.5mm}

        \includegraphics[
            width=0.90\linewidth
        ]{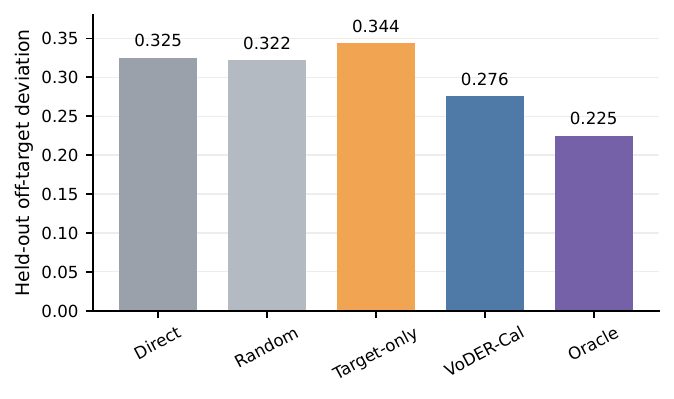}
    \end{minipage}

    \caption{
    Automatic evaluation of \vodercal{} at $B=3$ and $\rho=0.75$.
    \textbf{(a)} Candidates must retain a sufficient target response and pass content and speaker checks; \vodercal{} then selects the candidate with the smallest selector-side off-target deviation.
    \textbf{(b)} Percentage of requests satisfying the target, content, speaker, and held-out preservation criteria.
    \textbf{(c)} Held-out off-target deviation of selected candidates, where lower values indicate better preservation.
    }
    \label{fig:voder}
\end{figure*}

Figure~\ref{fig:voder} and Table~\ref{tab:voder} summarize the results. The two single-sample policies achieve joint success on approximately 4.8\% of requests. Expanding the pool to three candidates raises this rate to 14.1\% for Target-only, 14.3\% for \vodercal{}, and 14.6\% for the Oracle. Thus, most of the increase in binary success comes from candidate availability rather than the ranking rule itself: the paired difference between \vodercal{} and Target-only is $+0.19$ percentage points (95\% CI $[-0.56,0.93]$). Relative to single-sample policies, the corresponding differences are $+9.44$ points over Direct (95\% CI $[6.85,12.04]$) and $+9.53$ points over Random ($[7.53,11.60]$). Among accepted outputs, \vodercal{} retains 99.1\% of the strongest available target score on average.

\begin{table}[t]
\centering
\caption{Automatic results at $B=3$ and $\rho=0.75$, macro-averaged over the nine system--descriptor settings. Direct and Random use one descriptor-conditioned generation; Target-only, \vodercal{}, and Oracle select from the same three-candidate pool. ``Eligible'' denotes requests containing a target-, content-, and speaker-feasible candidate. ``All checks'' denotes joint success and additionally requires held-out off-target preservation.}
\label{tab:voder}
\resizebox{\columnwidth}{!}{\begin{tabular}{lrrrrr}
\toprule
Policy & Eligible & Target retain. & All checks & Held-out dev. & Calls\\
\midrule
Direct & 11.5\% & 99.0\% & 4.8\% & 0.325 & 1.0\\
Random & 11.2\% & 98.9\% & 4.7\% & 0.322 & 1.0\\
Target-only & 31.3\% & 99.4\% & 14.1\% & 0.344 & 3.0\\
VoDER-Cal & 31.3\% & 99.1\% & 14.3\% & 0.276 & 3.0\\
Oracle & 31.3\% & 99.1\% & 14.6\% & 0.225 & 3.0\\
\bottomrule
\end{tabular}
}
\end{table}

The matched-budget comparison is clearer in the continuous preservation measure. Mean held-out off-target deviation decreases from 0.344 for Target-only selection to 0.276 for \vodercal{}, compared with 0.225 for the Oracle. The reduction remains consistent for held-out speakers, held-out texts, and jointly held-out speaker--text combinations. \vodercal{} therefore improves preservation within the same three-candidate budget, even though its binary joint-success rate is statistically indistinguishable from Target-only selection.

\section{Auxiliary Perceptual Check}
\label{sec:human}

We conduct two blinded listening studies with ten adult listeners. H1 evaluates whether target and off-target changes are perceptible relative to a matched baseline, while H2 compares the candidates selected by \vodercal{} and Target-only selection. Each study contains 45 samples balanced across the three systems and three primary descriptors, with three independent ratings for each sample--task pair. Additional task and analysis details are provided in the supplementary material.

In H1, listeners compare each descriptor-conditioned output with its matched baseline and evaluate target and non-target changes separately. At the request level, listeners judge the requested attribute to have changed in 51.1\% of the 45 evaluated pairs (95\% CI: 37.8--64.4\%) and report a noticeable non-target change in 51.1\% (37.8--66.7\%). The two judgments partially overlap, with both target and non-target changes reported in 22.2\% of samples (11.1--35.6\%). These results provide perceptual evidence that both target responses and accompanying off-target changes can be noticeable.

\begin{figure*}[t]
\centering
\includegraphics[width=0.99\textwidth]{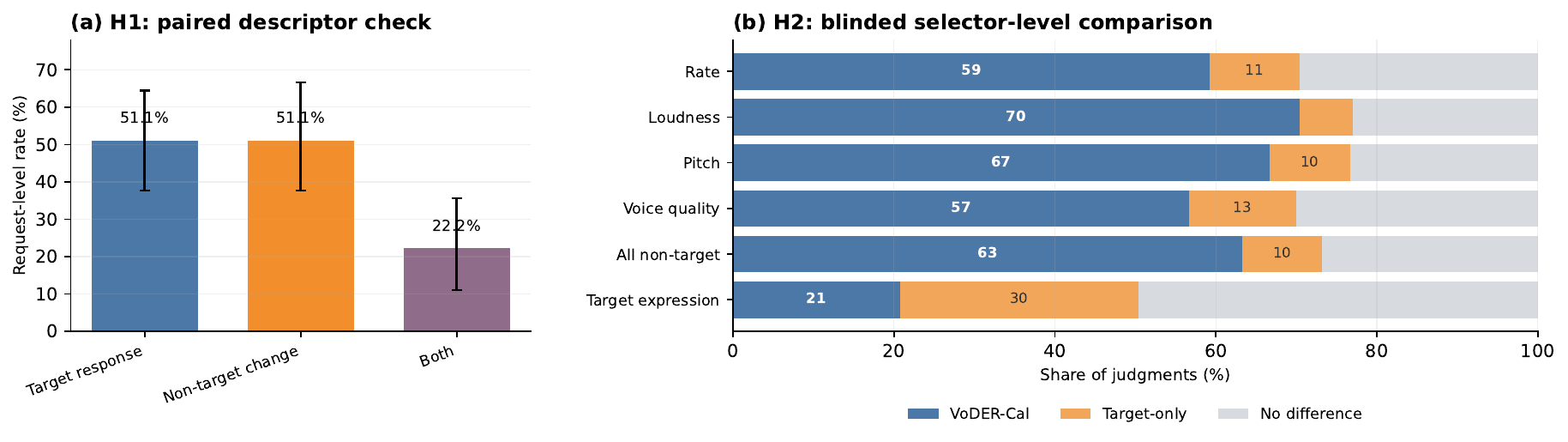}
\caption{
Perceptual evaluation.
\textbf{(a)} H1 compares descriptor-conditioned outputs with their matched baselines and separately evaluates target and off-target changes.
\textbf{(b)} H2 compares the candidates selected by \vodercal{} and Target-only selection using the matched baseline as reference. Error bars show request-clustered 95\% confidence intervals.
}
\label{fig:human}
\end{figure*}

In H2, listeners compare the candidates selected by \vodercal{} and Target-only selection, using the matched neutral output as reference. Across the four evaluated off-target dimensions, 63.3\% of all judgments favor the \vodercal{} candidate as closer to the baseline, 9.9\% favor the Target-only candidate, and 26.8\% report no noticeable difference. Among non-tied judgments, 86.4\% favor \vodercal{}. The preservation advantage is consistent across speaking rate, loudness, off-target pitch, and voice quality (Fig.~\ref{fig:human}).

Target-expression judgments show a different pattern: nearly half report no noticeable difference, while the remaining judgments more often favor Target-only selection. Automatic target-score retention therefore does not imply perceptually identical target strength, indicating a trade-off between stronger expression and tighter preservation.

\FloatBarrier

\section{Discussion}
\label{sec:discussion}

\subsection{Prompt adherence and attribute preservation are complementary}

Prompt adherence and attribute preservation capture different aspects of controllability. Prompt-adherence evaluation asks whether generated speech expresses the requested concept, whereas preservation-sensitive evaluation asks whether characteristics outside that concept remain stable. A system may perform well on the first criterion while changing several unrequested acoustic or prosodic dimensions. Evaluating both target response and off-target movement therefore provides a more complete account of natural-language voice control.

This distinction has direct implications for benchmark design. Descriptor-conditioned outputs should be compared with matched neutral baselines, and target response should be reported together with content validity, speaker preservation, and off-target change. Such an evaluation does not replace perceptual prompt-matching measures; it reveals a complementary property that global style preference or semantic agreement alone cannot capture.

The system-dependent response profiles further show that the same descriptor does not correspond to a consistent intervention across models. A request such as \emph{deep} may be realized mainly through lower pitch in one system, through timing and energy changes in another, or through broader spectral movement in a third. This variation may reflect correlations among voice characteristics in training data, the use of global style representations, or the underdetermined mapping from language to speech. Although the present output-level analysis does not identify the internal cause, it demonstrates the need for models that express descriptors through more predictable and controllable changes.

\subsection{VoDER-Cal and preservation-aware generation}

The candidate analysis shows that off-target coupling is not identical across samples generated for the same request. Candidates with similar target responses can differ substantially in how well they preserve other characteristics. \vodercal{} exploits this variation by selecting target-valid candidates with lower off-target deviation, providing a training-free way to improve control precision at inference time.

Under the current three-candidate pool, \vodercal{} recovers much of the preservation improvement measured by the held-out acoustic feature set. Because selector and evaluator features belong to the same correlated measurement family, the comparison should not be read as evidence of a perceptually optimal ranking rule. Further progress requires both more useful candidate variation and preservation signals that better track listener judgments.

Several model-level directions follow from these findings. Training objectives could jointly reward target expression and penalize off-target movement. Counterfactual speech pairs could provide direct supervision for what should change and what should remain stable. Structured representations may help separate pitch, timing, energy, spectral color, voice quality, and speaker identity. Finally, perceptually trained preservation models could align automatic selection more closely with listener judgments, particularly when stronger target expression competes with tighter preservation.

\subsection{Limitations}

The empirical findings are based on three reference-conditioned speech-generation systems and three primary descriptors. They establish a consistent pattern within this evaluation scope but do not imply that all systems or voice concepts behave identically. Higher-level descriptors such as \emph{warm}, \emph{professional}, or \emph{emotional} may require richer perceptual definitions than the signal-level hypotheses used here.

The target and off-target feature sets are operational evaluation choices rather than complete perceptual decompositions of voice attributes. Several measurements are correlated, so the number of shifted features should not be interpreted as the number of independent attributes affected. Spectral flatness and zero-crossing rate are also limited proxies for roughness, motivating stronger perceptual or learned measures of voice quality.

The listening studies provide perceptual support for the automatic findings but use a limited number of listeners and requests. H2 evaluates candidate choices when both selectors return an output rather than the full abstaining inference procedure. Larger studies with crossed listener--request models would support broader population-level conclusions.

Finally, \vodercal{} operates on a fixed candidate pool and depends on measurable target and off-target feature definitions. Its effectiveness therefore depends on whether suitable candidates occur in the sampling distribution and whether the preservation measurements reflect human perception. Inference-time selection can recover cleaner outputs, but eliminating systematic attribute coupling will ultimately require preservation to be incorporated into model representations, training objectives, or generation procedures.

\section{Conclusion}
\label{sec:conclusion}

Prompt adherence alone does not fully characterize natural-language voice control. When attribute-specific editing is desired, evaluation must also test whether characteristics outside the requested target remain stable. Across three speech-generation systems, our paired audit shows that descriptor-aligned responses frequently co-occur with off-target acoustic and prosodic changes. This coupling persists even among outputs whose target response clearly exceeds baseline seed variation, and its form differs substantially across systems.

We further introduce \vodercal{}, a training-free candidate-selection method that retains strong target responses while favoring lower off-target deviation. By exploiting variation within a small candidate pool, it improves automatic and listener-rated preservation without retraining the generator. Speech-generation models and benchmarks should therefore treat target expression and attribute preservation as joint objectives rather than infer precise control from prompt adherence alone.

\balance
\bibliographystyle{IEEEbib}
\bibliography{references}

\clearpage
\twocolumn[{
\centering
{\Large\bfseries Supplementary Material\par}
\vspace{3pt}
{\large
Beyond Prompt Adherence: Auditing Attribute-Level Voice Control in Speech Generation
\par}
\vspace{6pt}
}]

\section{Reproducibility and Evaluation Scope}

\subsection{Systems and model identifiers}

Table~\ref{tab:models} lists the three systems, model identifiers, and control interfaces used for generation. Environment specifications, inference settings, evaluation configurations, and analysis scripts are provided in the accompanying code repository.

\begin{table}[H]
\centering
\caption{Systems, model identifiers, and control interfaces.}
\label{tab:models}
\resizebox{\columnwidth}{!}{%
\begin{tabular}{lll}
\toprule
System & Model identifier & Interface \\
\midrule
\cosy{} & \texttt{FunAudioLLM/Fun-CosyVoice3-0.5B-2512} & instruction + reference \\
\vox{} & \texttt{openbmb/VoxCPM2} & inline instruction + reference \\
\fish{} & \texttt{fishaudio/s2-pro} & native inline tag + reference \\
\bottomrule
\end{tabular}%
}
\end{table}
\FloatBarrier

\subsection{Generation matrix}

For each system, the evaluation matrix contains six reference speakers, ten English texts, three random seeds, and eleven conditions: one neutral baseline, six natural-language voice descriptors, and four controls. Across the three systems, this gives
\begin{equation}
3\times(6+4+1)\times6\times10\times3=5{,}940
\end{equation}
generated outputs, or 1,980 outputs per system.

The primary cross-model audit and candidate-selection analysis focus on \emph{deep}, \emph{bright}, and \emph{rough}, for which the evaluation protocol defines directly comparable signal-level target proxies. The remaining descriptors---\emph{warm}, \emph{young}, and \emph{professional}---and the identity-preservation control are retained as supplementary audit conditions and are not part of the primary claims.

Each descriptor-conditioned output is paired with a matched neutral baseline generated using the same system, reference speaker, text, and seed. The three systems retain their native prompting interfaces. \cosy{} and \vox{} receive their supported instruction forms together with the reference speech. \fish{} receives its native inline voice tag. Free-form metainstructions are not treated as valid evidence for a system that does not expose such an instruction channel.

\section{Measurements and Statistical Definitions}

\subsection{Acoustic and prosodic measurements}

The analysis uses F0, speaking rate, duration, RMS energy, spectral centroid, 85\% spectral roll-off, spectral flatness, and zero-crossing rate. F0 is converted to semitone change for the primary paired audit:
\begin{equation}
\Delta f_{\mathrm{st}}=12\log_2\frac{f(y_d)}{f(y_0)}.
\end{equation}
Table~\ref{tab:scales} gives the fixed scales used to display comparable normalized movement and to construct candidate-selection distances. A value of one represents one fixed scale, not a perceptual just-noticeable difference.

\begin{table}[H]
\centering
\caption{Fixed physical scales used for normalization.}
\label{tab:scales}
\begin{tabular}{lr}
\toprule
Feature & Fixed scale\\
\midrule
F0 & 3 semitones\\
Speaking rate & 0.8 words/s\\
Spectral centroid & 220 Hz\\
Spectral roll-off & 450 Hz\\
Spectral flatness & 0.03\\
Zero-crossing rate & 0.03\\
RMS energy & 0.04\\
Duration & 1.0 s\\
\bottomrule
\end{tabular}

\end{table}

The target feature sets are:
\begin{align}
\textit{deep}:&\quad \{\Delta\mathrm{F0},\Delta\mathrm{rolloff}\},\\
\textit{bright}:&\quad \{\Delta\mathrm{centroid},\Delta\mathrm{rolloff}\},\\
\textit{rough}:&\quad \{\Delta\mathrm{flatness},\Delta\mathrm{ZCR}\}.
\end{align}
For deep, lower values are target-aligned; for bright and rough, higher values are target-aligned. These sets define the scope of the signal-level audit rather than universal acoustic definitions of the descriptors. Flatness and zero-crossing rate capture noisy or spectrally diffuse signal behavior but do not constitute a complete perceptual roughness model; results for \emph{rough} should therefore be interpreted as signal-level evidence. Several measurements are correlated---particularly speaking rate with duration and centroid with roll-off---so shifted-feature counts describe response breadth rather than independent perceptual attributes.

\subsection{Target score and validity gates}

For descriptor $d$, let $S_d$ denote its target feature set, $a_j$ the fixed scale of feature $j$, and $s_{d,j}\in\{-1,+1\}$ its target direction. Candidate $i$ receives the direction-aligned score
\begin{equation}
T_i(d)=\frac{1}{|S_d|}\sum_{j\in S_d}s_{d,j}\frac{z_j(y_i)-z_j(y_0)}{a_j}.
\end{equation}
The same definition is used to estimate the one-sided baseline-noise threshold and to enforce target retention in \vodercal{}. The target score averages direction-aligned, normalized changes across the descriptor-specific target features. Candidate feasibility additionally requires the target score to exceed both the baseline-noise threshold and the retained fraction of the strongest available response, while passing the content and speaker checks.

Content validity is computed from normalized ASR transcripts, and speaker preservation from cosine similarity between generated and reference ECAPA-TDNN embeddings. Their binary thresholds are fixed on the calibration split and then applied unchanged to evaluation requests; x-vector similarity is used only as a supplementary cross-check. Model identifiers, transcript-normalization rules, calibration assignments, threshold values, and the target-noise quantile are specified in the evaluation configuration provided with the code. The same settings are applied unchanged to all evaluation requests.

\subsection{Request-level inference}

A request block is a unique system--descriptor--speaker--text combination. Seed-level deltas are averaged within a request, and bootstrap samples resample request blocks with replacement. This treats the three seeds as repeated generations rather than independent experimental units. Unless otherwise noted, confidence intervals use 10,000 replicates.

For the conditional analysis, target-noise thresholds are obtained from the one-sided target-direction distribution of paired baseline seeds. An output is target-responsive when its target score exceeds this threshold. An off-target deviation is called substantial when its absolute normalized magnitude reaches 0.5. The output-level fraction is accompanied by request-clustered uncertainty so that multiple seeds from the same request are not treated as independent. The threshold is used to summarize frequency; the main heatmap and feature-level confidence intervals retain the continuous measurements.

\section{Detailed Audit Results}

\subsection{Conditional target-response analysis}

Table~\ref{tab:conditional} gives the complete conditional summary. In eight sufficiently populated settings, 54.5\%--95.8\% of target-responsive outputs exhibit at least one substantial off-target deviation. The \fish{} deep setting contains only two target-responsive outputs under the composite threshold and is therefore descriptive. This is compatible with the simpler F0-direction result in the main audit because the composite criterion requires a joint response across both target features.

\begin{table}[H]
\centering
\caption{Conditional audit. ``Responsive'' counts outputs exceeding the target-noise threshold; ``Requests'' counts represented speaker--text blocks. ``Shifted features'' counts off-target feature intervals excluding zero.}
\label{tab:conditional}
\resizebox{\columnwidth}{!}{\begin{tabular}{llrrrr}
\toprule
Model & Desc. & Responsive & Requests & $\geq$1 off-target & Shifted features\\
\midrule
CosyVoice3 & Deep & 32/180 & 25 & 93.8\% & 4\\
 & Bright & 99/180 & 52 & 89.9\% & 5\\
 & Rough & 57/180 & 42 & 87.7\% & 5\\
\addlinespace[1pt]
VoxCPM2 & Deep & 24/180 & 19 & 95.8\% & 2\\
 & Bright & 55/180 & 39 & 87.3\% & 3\\
 & Rough & 51/180 & 36 & 94.1\% & 3\\
\addlinespace[1pt]
Fish-Speech-S2 & Deep & 2/180$^{\ddagger}$ & 2 & 100.0\% & 6\\
 & Bright & 55/180 & 44 & 54.5\% & 2\\
 & Rough & 53/180 & 41 & 81.1\% & 3\\
\bottomrule
\end{tabular}
}
\begin{flushleft}\footnotesize
$^{\ddagger}$ Fewer than ten responsive outputs; descriptive only.
\end{flushleft}
\end{table}

\subsection{Control conditions}

Table~\ref{tab:controls} summarizes the positive and generic instruction controls. The explicit lower-pitch condition decreases mean F0 in every system, verifying the directionality of the interface and measurement pipeline. Neutral and nonsense conditions can also shift several features, particularly in \cosy{} and \vox{}. These results show that additional prompting can itself introduce acoustic variation, reinforcing the value of comparing descriptor-conditioned outputs with matched neutral baselines.

\begin{table}[H]
\centering
\caption{Control summary. F0 values are request-block mean changes in Hz with 95\% bootstrap intervals. ``Shifted'' counts the eight measurements whose interval excludes zero.}
\label{tab:controls}
\resizebox{\columnwidth}{!}{\begin{tabular}{lrrr}
\toprule
Model & Lower-pitch F0 (Hz) & Neutral shifted & Nonsense shifted\\
\midrule
CosyVoice3 & -7.33 [-9.45,-5.30] & 5/8 & 5/8\\
VoxCPM2 & -7.17 [-12.38,-1.77] & 7/8 & 5/8\\
Fish-Speech-S2 & -4.80 [-6.43,-3.12] & 0/8 & 2/8\\
\bottomrule
\end{tabular}
}
\end{table}

All control analyses use the complete eight-feature measurement set, including spectral flatness and zero-crossing rate. Roughness-related control results remain signal-level diagnostics because these measurements provide limited perceptual proxies for roughness.

\section{VoDER-Cal Details and Robustness}

\subsection{Policies and operating point}

The candidate budget counts descriptor-conditioned generations; the matched neutral baseline $y_0$ is shared across all policies and is not included in $B$. Direct uses the descriptor-conditioned candidate generated with the fixed first seed. Random draws one seed and returns the corresponding candidate without inspecting the remaining candidates. Target-only, \vodercal{}, and the Oracle inspect the same three-seed candidate pool.

Target-only selects the feasible candidate with the strongest target response. \vodercal{} selects the feasible candidate with the smallest selector-side off-target deviation. The Oracle selects the candidate with the smallest held-out off-target deviation while remaining restricted to the same three candidates.

The main operating point uses $B=3$ and $\rho=0.75$. Eligibility depends only on the target, content, and speaker criteria. The joint-success outcome additionally requires held-out off-target preservation. Requests without a feasible candidate are counted as failures in the all-request rate. The held-out preservation threshold is applied only when evaluating the selected output, while eligibility is determined by the target, content, and speaker criteria.

\subsection{Paired comparisons}

Table~\ref{tab:bootstrap} reports request-level paired bootstrap differences in joint success. The large gains over Direct and Random primarily reflect the availability of three candidates. The matched-budget difference between \vodercal{} and Target-only is small and its interval includes zero; continuous held-out deviation and the listening comparison provide the stronger evidence for the preservation-aware ranking.

\begin{table}[H]
\centering
\caption{Macro paired differences in joint-success rate, in percentage points, at $B=3$ and $\rho=0.75$.}
\label{tab:bootstrap}
\begin{tabular}{lcc}
\toprule
Comparison & Difference & 95\% CI \\
\midrule
\vodercal{} $-$ Direct & $+9.44$ & $[+6.85,+12.04]$ \\
\vodercal{} $-$ Random & $+9.53$ & $[+7.53,+11.60]$ \\
\vodercal{} $-$ Target-only & $+0.19$ & $[-0.56,+0.93]$ \\
Oracle $-$ \vodercal{} & $+0.37$ & $[+0.00,+0.93]$ \\
\bottomrule
\end{tabular}
\end{table}

\subsection{Held-out speakers and texts}

Table~\ref{tab:oos} reports macro results when thresholds and selection definitions are applied to unseen speakers, unseen texts, or both. \vodercal{} maintains lower continuous held-out deviation than Target-only selection and remains close to the candidate-pool Oracle.

\begin{table}[H]
\centering
\caption{Strict out-of-sample results at the main operating point.}
\label{tab:oos}
\resizebox{\columnwidth}{!}{\begin{tabular}{llrrr}
\toprule
Split & Policy & Eligible & All checks & Held-out dev.\\
\midrule
Unseen speaker & Direct & 12.6\% & 4.6\% & 0.325\\
 & Target-only & 32.8\% & 14.1\% & 0.344\\
 & VoDER-Cal & 32.8\% & 14.1\% & 0.277\\
 & Oracle & 32.8\% & 14.4\% & 0.227\\
\addlinespace[1pt]
Unseen text & Direct & 13.1\% & 5.6\% & 0.325\\
 & Target-only & 34.4\% & 16.5\% & 0.344\\
 & VoDER-Cal & 34.4\% & 16.9\% & 0.276\\
 & Oracle & 34.4\% & 17.2\% & 0.228\\
\addlinespace[1pt]
Unseen speaker+text & Direct & 13.3\% & 4.6\% & 0.325\\
 & Target-only & 34.4\% & 15.4\% & 0.343\\
 & VoDER-Cal & 34.4\% & 15.7\% & 0.277\\
 & Oracle & 34.4\% & 16.1\% & 0.228\\
\bottomrule
\end{tabular}
}
\end{table}

\subsection{Target-retention sensitivity}

Table~\ref{tab:rho} varies the target-retention fraction while keeping the rest of the method fixed. Lower $\rho$ admits more requests and permits lower off-target movement, but the selected output can retain less of the strongest available target response. The main value $\rho=0.75$ retains 99.1\% of the strongest target score among accepted outputs.

\begin{table}[H]
\centering
\caption{Sensitivity to the target-retention fraction $\rho$.}
\label{tab:rho}
\resizebox{\columnwidth}{!}{\begin{tabular}{rrrrr}
\toprule
$\rho$ & Eligible & Target retain. & All checks & Held-out dev.\\
\midrule
0.00 & 34.4\% & 87.9\% & 17.4\% & 0.269\\
0.50 & 32.4\% & 95.2\% & 15.4\% & 0.272\\
0.75 & 31.3\% & 99.1\% & 14.3\% & 0.276\\
0.90 & 30.4\% & 99.9\% & 13.9\% & 0.276\\
\bottomrule
\end{tabular}
}
\end{table}

We also evaluate balanced selector/evaluator feature partitions, alternative preservation quantiles, and calibration-only robust scaling. The gains over Direct and Random persist across these analyses, and \vodercal{} maintains lower continuous held-out deviation than Target-only across the reported out-of-sample settings. Thresholds, feature definitions, and the three-candidate budget are shared across systems.

\section{Listening-Study Details}

\subsection{Design}

Ten adult listeners completed two blinded studies. H1 used 45 matched-baseline--descriptor pairs balanced across three systems, three descriptors, and five requests per system--descriptor cell. H2 used 45 matched-baseline--candidate triplets with Target-only and \vodercal{} identities hidden and positions balanced. Each sample--task pair received three independent responses.

The tasks were decomposed into direct perceptual judgments. H1 separately assessed the requested attribute and non-target dimensions. For H1, the three responses assigned to each request--task pair are combined by majority judgment, and the resulting outcomes are summarized across the 45 evaluated pairs. H2 separately compared target expression and closeness to the matched baseline in speaking rate, loudness, non-target pitch, voice quality, speaker identity, and naturalness. The H2 study is a selector-level comparison: it asks which candidate is preferred when both policies return a candidate, rather than evaluating the complete abstaining pipeline.

\subsection{Clustered outcomes}

Table~\ref{tab:humanfull} reports all H2 endpoints with both request-clustered and listener-clustered intervals. The aggregate four-dimension off-target result contains 540 individual judgments: 45 requests, four off-target dimensions, and three independent responses per request--dimension pair. Request-clustered intervals resample requests while retaining all associated dimensions and responses; listener-clustered intervals analogously resample listeners. These complementary intervals account for each dependence source separately, and the non-target preference is consistent across clustering choices. The 86.4\% non-tied value in the main manuscript is computed from the 63.3\% \vodercal{} and 9.9\% Target-only shares; the 26.8\% ``no difference'' responses remain visible in the full-distribution figures and tables.

\begin{table}[H]
\centering
\caption{Full H2 selector-level perceptual results. Percentages include ties. ``Non-tied \vodercal{}'' conditions on a preference for either selector.}
\label{tab:humanfull}
\resizebox{\columnwidth}{!}{%
\begin{tabular}{lcccccc}
\toprule
Endpoint & \vodercal{} & Target-only & Tie & Request 95\% CI & Listener 95\% CI & Non-tied \\
\midrule
Rate & 59.3\% & 11.1\% & 29.6\% & [51.1,66.7] & [52.7,66.3] & 84.2\% \\
Loudness & 70.4\% & 6.7\% & 23.0\% & [62.2,77.8] & [60.9,80.0] & 91.3\% \\
Pitch & 66.7\% & 10.0\% & 23.3\% & [55.6,76.7] & [55.5,76.4] & 87.0\% \\
Voice quality & 56.7\% & 13.3\% & 30.0\% & [50.0,63.3] & [48.7,63.7] & 81.0\% \\
Combined & 63.3\% & 9.9\% & 26.8\% & [59.2,67.2] & [59.4,67.3] & 86.4\% \\
Target expression & 20.7\% & 29.6\% & 49.6\% & [14.8,26.7] & [13.9,28.4] & 41.2\% \\
Speaker similarity & 33.3\% & 35.6\% & 31.1\% & [25.9,41.5] & [27.3,40.6] & 48.4\% \\
Naturalness & 39.3\% & 34.1\% & 26.7\% & [30.4,48.1] & [36.0,42.3] & 53.5\% \\
\bottomrule
\end{tabular}
}
\end{table}

\subsection{Agreement and sensitivity}

Table~\ref{tab:agreement} summarizes pairwise exact agreement. H1 five-point judgments also report agreement within one scale point. Fine-grained voice judgments show moderate variability, motivating the use of clustered intervals and separate perceptual dimensions. The strong H2 non-target preference is not explained by a fixed A/B position or a particular trial-order segment (Table~\ref{tab:position}).

\begin{table}[H]
\centering
\caption{Pairwise inter-rater agreement.}
\label{tab:agreement}
\resizebox{\columnwidth}{!}{\begin{tabular}{llrr}
\toprule
Stage & Dimension & Exact agreement & Within-one\\
\midrule
H1 & loudness & 22.2\% & 71.1\%\\
H1 & naturalness & 31.9\% & 76.3\%\\
H1 & pitch & 22.2\% & 64.4\%\\
H1 & speaker similarity & 39.3\% & 84.4\%\\
H1 & speaking rate & 22.2\% & 64.4\%\\
H1 & target attribute & 40.7\% & 80.7\%\\
H1 & voice quality & 37.8\% & 76.7\%\\
H2 & loudness & 54.1\% & --\\
H2 & naturalness & 31.9\% & --\\
H2 & pitch & 53.3\% & --\\
H2 & speaker similarity & 34.1\% & --\\
H2 & speaking rate & 42.2\% & --\\
H2 & target attribute & 34.8\% & --\\
H2 & voice quality & 33.3\% & --\\
\bottomrule
\end{tabular}
}
\end{table}

\begin{table}[H]
\centering
\caption{H2 sensitivity to candidate position and trial order. Percentages include ties.}
\label{tab:position}
\resizebox{\columnwidth}{!}{%
\begin{tabular}{lllccc}
\toprule
Endpoint & Factor & Level & \vodercal{} & Target-only & Tie \\
\midrule
\multirow{5}{*}{Non-target preservation}
 & Candidate position & A & 64.0\% & 10.8\% & 25.3\% \\
 & Candidate position & B & 63.3\% & 9.5\% & 27.3\% \\
 & Trial order & Early & 64.9\% & 11.7\% & 23.4\% \\
 & Trial order & Middle & 61.0\% & 12.3\% & 26.7\% \\
 & Trial order & Late & 64.7\% & 6.0\% & 29.3\% \\
\midrule
\multirow{5}{*}{Target expression}
 & Candidate position & A & 17.5\% & 29.8\% & 52.6\% \\
 & Candidate position & B & 23.1\% & 29.5\% & 47.4\% \\
 & Trial order & Early & 20.0\% & 33.3\% & 46.7\% \\
 & Trial order & Middle & 22.2\% & 28.9\% & 48.9\% \\
 & Trial order & Late & 20.0\% & 26.7\% & 53.3\% \\
\bottomrule
\end{tabular}%
}
\end{table}

\FloatBarrier
\section{Evidence Scope}

The empirical conclusion is behavioral and limited to the tested systems and signal-level target definitions: descriptor-aligned outputs can also change acoustic and prosodic measurements outside the corresponding target set. \vodercal{} shows that preservation-aware ranking can select candidates with lower measured off-target deviation from a fixed pool. Reducing systematic coupling at its source remains a model-training and representation-learning problem.

\end{document}